\documentclass[journal=jacsat,manuscript=article]{achemso}

\usepackage[version=3]{mhchem} 
\usepackage{xcolor}
\usepackage{booktabs}
\usepackage[numbers]{natbib}  
\usepackage{xcolor}            
\usepackage{graphicx}          
\usepackage{float}
\usepackage{subcaption}
\usepackage[normalem]{ulem} 
\usepackage{graphicx,lipsum}  
\usepackage{subcaption,xcolor}  
\usepackage[colorlinks=true, allcolors=blue]{hyperref}

\title{Geometrical Imperfections in a Digital Quadrupole Mass Filter: A Comprehensive Simulation Study in the First Stability Zone}

\author{Brotin Taraphdar}
\affiliation{School of Chemical Sciences, Indian Association for the Cultivation of Science, Kolkata-700032, India}

\author{Sukanya Jana}
\affiliation{School of Chemical Sciences, Indian Association for the Cultivation of Science, Kolkata-700032, India}

\author{Pintu Mandal}
\affiliation{Department of Physics, St. Paul\text{'}s Cathedral Mission College, Kolkata-700009, India}
\email{pintuphys@gmail.com}

\author{Nabanita Deb}
\affiliation{School of Chemical Sciences, Indian Association for the Cultivation of Science, Kolkata-700032, India}
\email{nabanita.deb@iacs.res.in}

\begin{document}

\begin{abstract}
Geometrical imperfections in quadrupole mass filters introduce higher-order field components that can significantly influence device performance, particularly under non-sinusoidal excitation. In this work, a comprehensive simulation study is carried out to investigate the effect of geometrical imperfections on the performance of a rectangular wave driven quadrupole mass filter operating in the first stability zone. Radial field distortions arising from controlled variations in rod geometry and position, including single rod radius variation, single rod displacement, diagonal rod radius variation, and diagonal rod displacement, are examined. These imperfections introduce octupole field components that distort the ideal quadrupolar field distribution. The influence of such distortions on key performance parameters, namely mass resolution and ion transmission efficiency, is systematically evaluated. The results show that the presence of radial asymmetry leads to a degradation of both resolution and transmission efficiency in all cases considered. Furthermore, the study reveals a strong dependence of mass filter performance on the initial state of the applied pulsed waveform, specifically whether the asymmetric rod pair is subjected to the high or low level of the RF pulse.  These findings provide important insights into the tolerance limits of geometrical imperfections and their impact on the performance of pulsed wave driven quadrupole mass filters, which are relevant for the design and optimization of high-resolution digital mass filtering systems. 
\end{abstract}

\textbf{KEYWORDS}\newline
digital quadrupole mass filter, rectangular wave excitation, geometrical imperfections, octupole fields, transmission characteristics

\section{Introduction}

The quadrupole mass filter (QMF) has conventionally operated using sinusoidal RF and DC voltages, where ion stability is governed by solutions of the Mathieu equation \cite{paul1953neues,paul1955elektrische}. In practical realizations employing circular electrodes, the ideal quadrupolar field is accompanied by higher-order multipole components, notably the dodecapole term~\cite{dawson2013quadrupole}. Additional geometric imperfections, such as electrode misalignment or radius variation, introduce further distortions, including octupole components, which are known to influence the performance of sinusoidally driven QMFs.~\cite{bugrov2023modelling, douglas2002influence,konenkov2006linear,sysoev2022balance,taylor2008prediction,ding2003quadrupole,bugrov2023simulation,mandal2024non,douglas1999spatial}
Recent studies from our group have shown that even small structural asymmetries, introduced via single-electrode radius imperfections, lead to the emergence of higher-order multipole components, particularly the octupole field, resulting in reduced resolution in a linear QMF operating in the first stability zone under sinusoidal excitation~\cite{Jana2025}. Similar effects are observed for diagonal radial asymmetry introduced through electrode displacement or radius variation, with the exception of a specific displacement condition that yields enhanced resolution~\cite{sysoev2022balance,Jana2026}. The influence of octupole field has also been extended to the second stability zone in sinusoidally driven QMFs~\cite{dutta2026influence}. These studies establish that geometric asymmetry significantly alters QMF performance under sinusoidal excitation through the introduction of higher-order multipole fields.

Digital (rectangular) waveform excitation offers an alternative operational paradigm in which ion stability is governed by timing parameters rather than analog voltage ratios. This concept, originally proposed in early work~\cite{Richards1973}, has matured into a well-established approach in recent years~\cite{simke2022simulations,hu2024simulation,Schrader2024,ding2026development}. The use of digital waveform relaxes electronic constraints, as precise timing control is generally easier to achieve than maintaining stable RF/DC voltage ratios. In such systems, mass selection is achieved through frequency variation at fixed amplitude, reducing the requirement for large RF voltages and extending the accessible $m/z$ range~\cite{Capek24}. Furthermore, the duty cycle serves as an additional control parameter, enabling modulation of stability regions and facilitating ion transmission, trapping, and ejection within a unified framework~\cite{brabeck2014mapping}.

In contrast to sinusoidal excitation, a rectangular waveform exhibits a richer spectral content, consisting of a DC component for duty cycles other than $50\%$ and harmonics at integer multiples of the fundamental frequency. 
This intrinsic DC contribution, together with higher harmonics, modifies the effective potential experienced by ions and alters the mapping between applied voltages and the $a$--$q$ stability space\cite{bandelow2013stability,ding2026development}. Consequently, the interplay between waveform structure and geometric asymmetry is expected to produce stability behavior and transmission characteristics that differ qualitatively from those under sinusoidal excitation. Despite these distinctive features and the additional control parameters available under digital excitation, the transmission characteristics in the presence of geometrical imperfections remain insufficiently understood.

\begin{figure}
    \centering
    \includegraphics[width=1\linewidth]{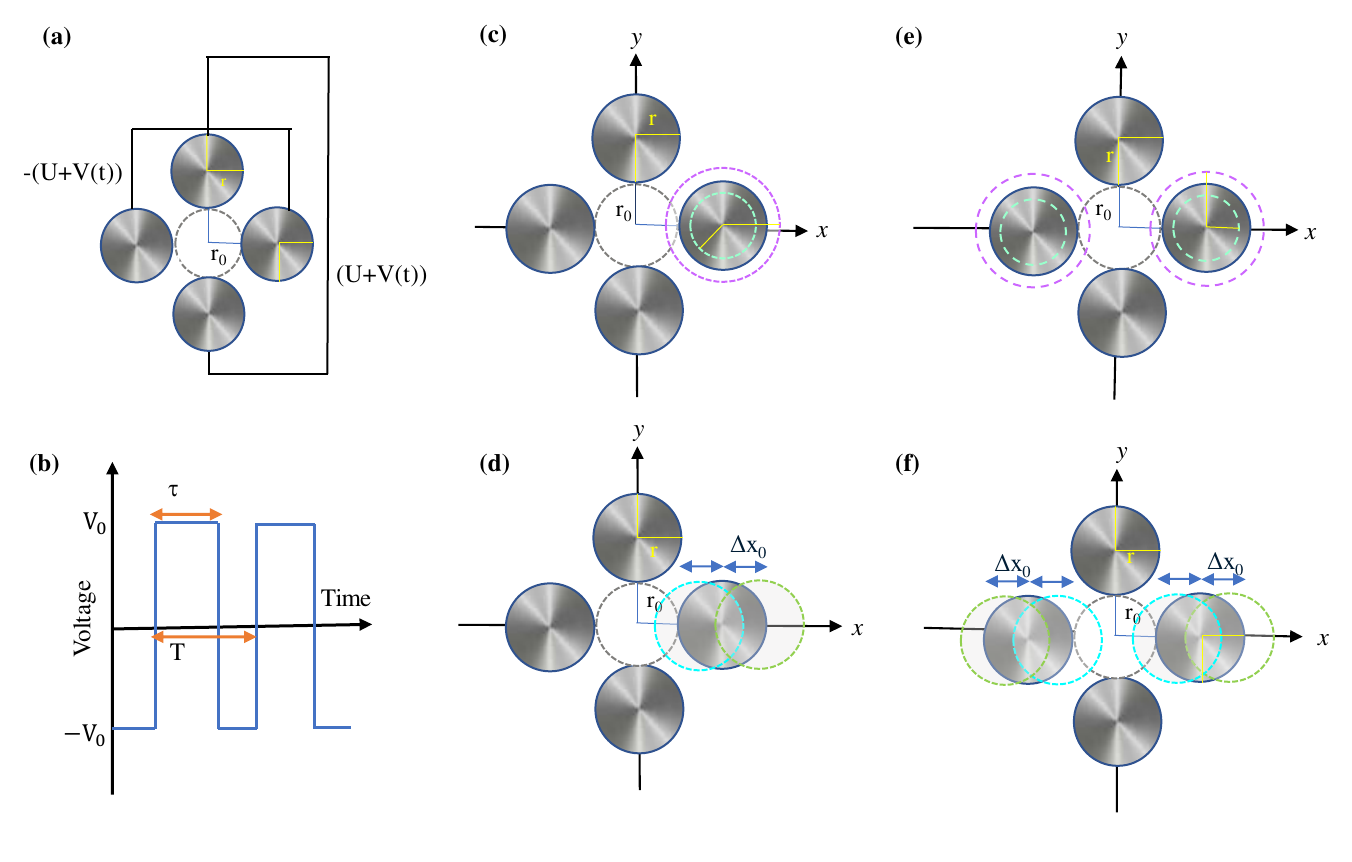}
    \caption{Schematic representation of the QMF cross-section in the $x$--$y$ radial plane: (a) geometrically symmetric QMF with electrical connections; (b) typical rectangular excitation pulse; (c) single-electrode radius variation asymmetry; (d) single-electrode displacement asymmetry; (e) diagonal asymmetry introduced through radius variation; and (f) diagonal asymmetry introduced through electrode displacement.}
    \label{fig:scheme}
\end{figure}

In the present work, we investigate the effects of controlled geometric asymmetry in a digitally driven quadrupole mass filter operating in the first stability zone. Radial asymmetry is introduced through single-rod perturbations (radius variation and displacement) as well as diagonal rod-pair modifications, as illustrated in Fig.~\ref{fig:scheme}. The resulting stability behavior, transmission characteristics, and resolution are systematically analyzed and compared with the symmetric configuration. This approach isolates the impact of radial field distortions on QMF performance under rectangular waveform excitation. 

The QMF 
geometry employed in this study is shown schematically in Fig.~\ref{fig:scheme}. A quadrupole mass filter of length $200\,\mathrm{mm}$ and field radius $r_0 = 4\,\mathrm{mm}$ is considered, with a $1\,\mathrm{mm}$ gap between the ion source and the QMF entrance. The ions are initialized with a longitudinal kinetic energy of $10\,\mathrm{eV}$ and uniformly distributed initial positions over a circular region of diameter $0.2\,\mathrm{mm}$. The ion birth times are uniformly distributed within the interval $0$--$0.63\,\mu\mathrm{s}$. All simulations are carried out for $\mathrm{Ca}^+$ ions with $m/z = 40$ \cite{yang2023simulation}.

\subsection{Theoretical aspects of Digital QMF}
Unlike a conventional QMF, which operates by varying both the DC voltage and the RF amplitude, a digital QMF achieves mass filtering by adjusting the drive frequency at a fixed amplitude and duty cycle, $d=\tau/T$ where $\tau$ is the duration of the high state and $T$ is the period of excitation. The rectangular waveform of amplitude $V_0$ is defined as $V(t)=V_0 S(t)$ where,
\begin{equation}
    S(t) =
\begin{cases}
+1, & 0 \leq t < \tau \\
-1, & \tau \leq t < T
\end{cases}
\end{equation}
and $S(t+T)=S(t)$.

\textcolor{blue}{}

The general form of the electric potential inside the QMF is given by
\begin{equation}
    \Phi(x,y,t) = \sum_{N=0}^{\infty} \Phi_N(x,y)(U+V_0S(t))
\end{equation}
where, 
\begin{equation}
    \Phi_N(x,y) = Re\text{    } A_N \left(\frac{z}{r_0}\right)^N, z = x + iy.
\end{equation}
$A_N$ are the weighting coefficients corresponding to the multipole expansion and $r_0$ is the effective field radius as shown in Fig.~\ref{fig:scheme}. Pure quadrupole potential corresponds to $N=2$ and the general form of the potential takes the following form:
\begin{equation}
    \Phi(x,y,t)=\frac{\left(x^2-y^2\right)}{r_0^2}(U+V_0S(t)).
\end{equation}
The motion of an ion of charge $e$ and mass $m$ in the radial plane of the QMF is governed by the Hill differential equation:
\begin{equation}
    \frac{d^2u}{d\xi^2}+(a_u + 2q_u S(\xi))u = 0,
\end{equation}
where $u = x, y$ and $\xi=\omega t/2$ with $\omega=2\pi/T$. Dimensionless parameters $a_u$ and $q_u$ are defined as
\begin{equation}\label{eq7}
a_x = -a_y = \frac{8eU}{m r_{0}^2 \omega^2}, \quad 
q_x = -q_y = \frac{4eV_0}{m r_{0}^2 \omega^2}. 
\end{equation}

To obtain the stability criteria for ion motion, the Hill equation is solved, which is a generalized form of the Mathieu equation. This is done through a matrix-based method involving Floquet Theory for first-order differential equations. Therefore, the second-order Hill equation is transformed into an equivalent first-order matrix differential equation. The solution of this system is expressed in terms of the monodromy (or transfer) matrix, denoted by $M$.
The stability criterion is obtained from the properties of this matrix and is given by:
\begin{equation}
    |\mathrm{Tr}(M)| \leq 2, 
\end{equation}
where $\mathrm{Tr}(M)$ is the trace of the transfer matrix. This condition determines the stability of ion motion and allows the generation of stability diagrams for various duty cycles\cite{konenkov2002matrix,pipes1953matrix,march2005quadrupole,dawson2013quadrupole,brabeck2016computational,brabeck2016characterization}.

The introduction of an additional control parameter, the duty cycle ($d$), provides enhanced flexibility in tailoring the $a$--$q$ stability space and, consequently, the operational window of a digital QMF. In typical operation, the digital QMF is driven at $U = 0$ (i.e., $a = 0$), and mass filtering is achieved by scanning the RF frequency while maintaining a constant amplitude ($V_0 = \mathrm{const.}$)~\cite{hu2024simulation,ivanov2025second,schrader2023optimization}. Under this condition, the conventional $a$--$q$ stability diagram can be mapped onto an equivalent duty cycle--frequency stability representation.

For a fixed duty cycle, the condition $a=0$ intersects the stability region at two $q$ values, which map to the lower and upper cutoff frequencies ($q \propto 1/\omega^2$), thereby defining the stable frequency window. 
Variation of the duty cycle shifts these boundaries, forming the basis of the $d-f$ space stability diagram used to determine accessible operating regimes, as demonstrated earlier by Simke et al.~\cite{simke2022simulations}.

\section{Digital QMF: Symmetric Geometry}
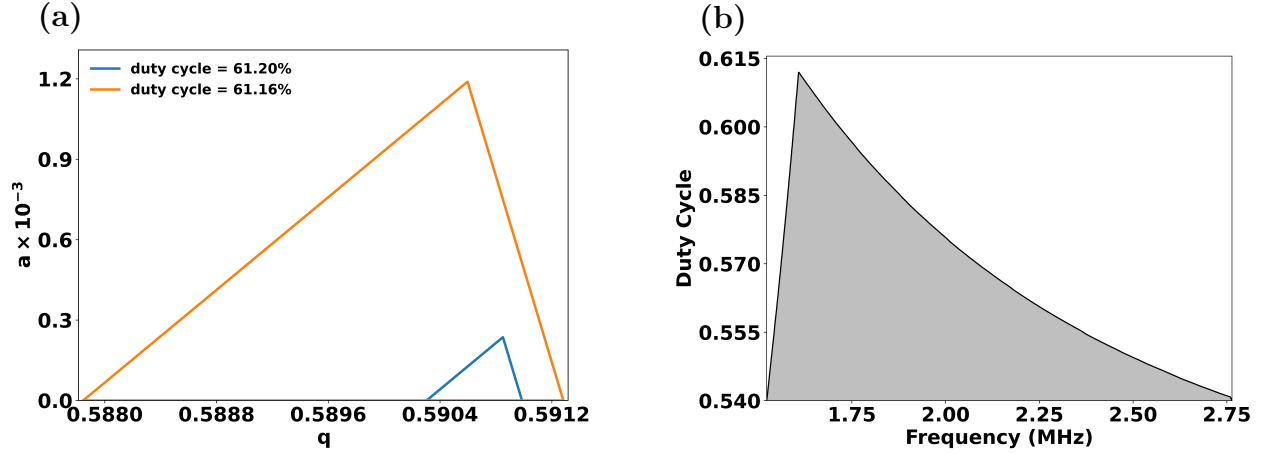
\begin{figure*}[htbp]
    \centering
    \begin{subfigure}[b]{0.47\textwidth}
        \centering
        \setlength{\unitlength}{1mm} 
        \begin{picture}(0,0)
            \put(-35, 2){\textbf{(a)}} 
        \end{picture}
        \includegraphics[width=\textwidth]{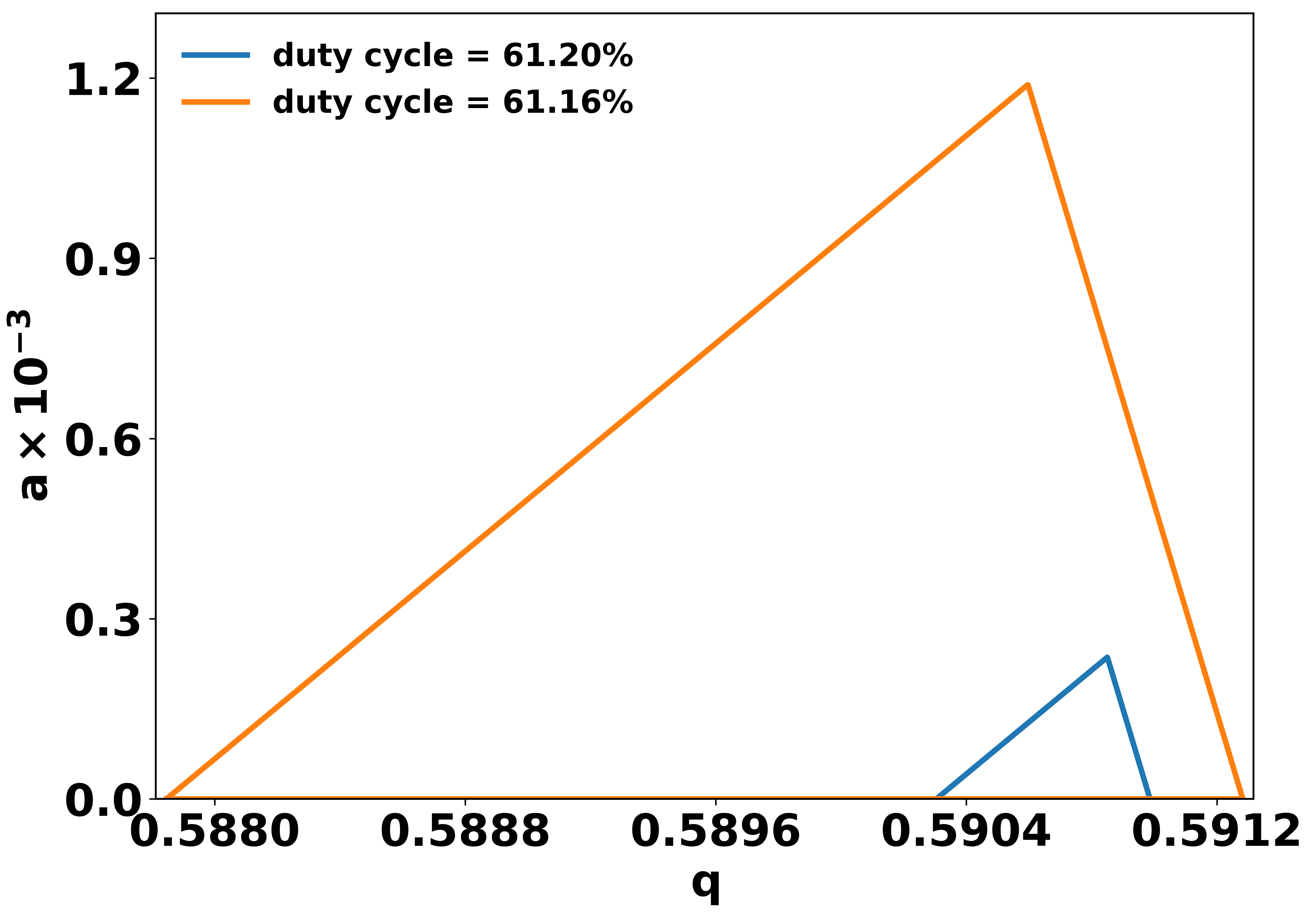}
        \label{fig:aq-duty cycle}
    \end{subfigure}
    \hfill
    \begin{subfigure}[b]{0.47\textwidth}
        \centering
        \setlength{\unitlength}{1mm}
        \begin{picture}(0,0)
            \put(-35, 2){\textbf{(b)}}
        \end{picture}
        \includegraphics[width=\textwidth]{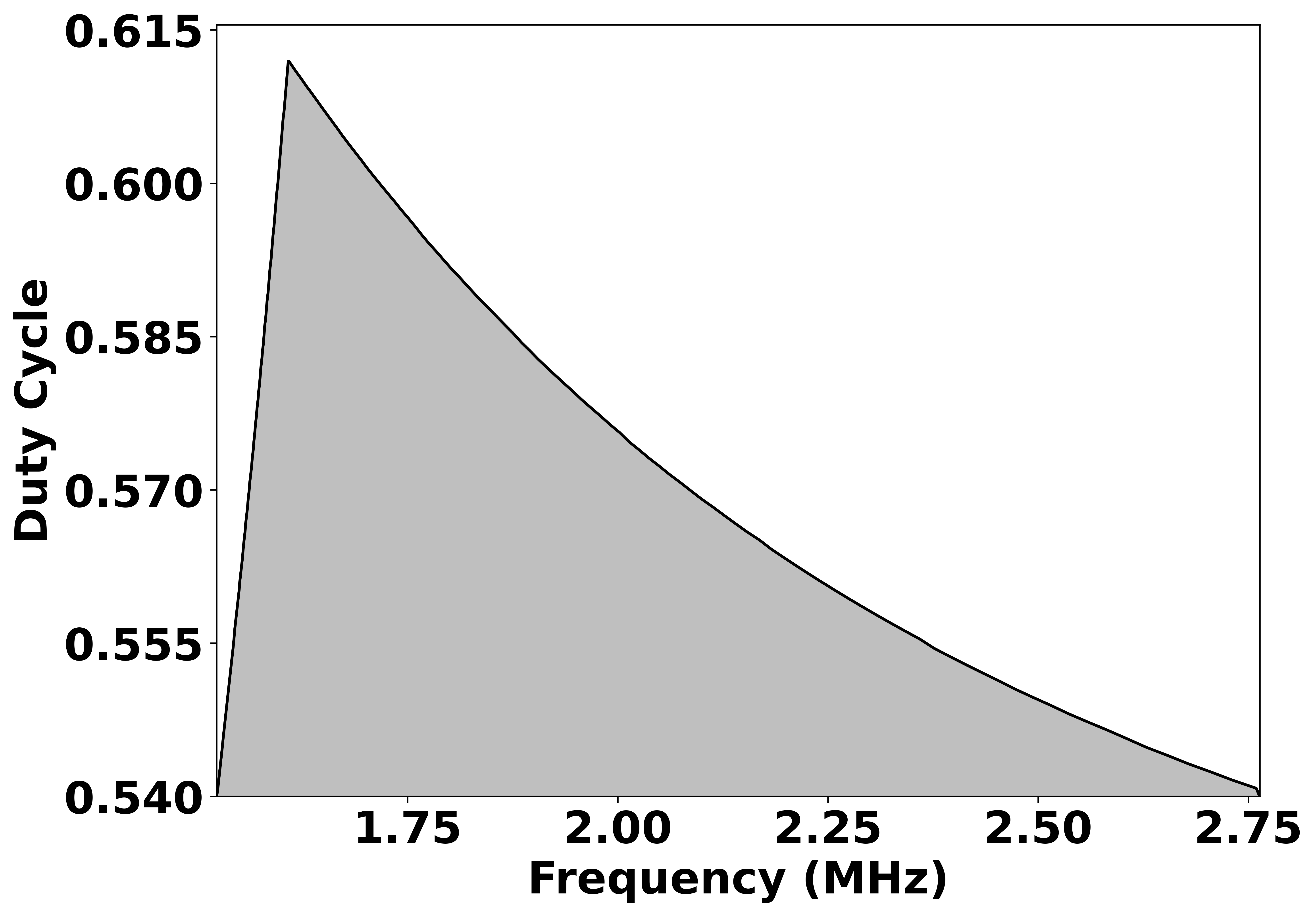}
        \label{fig:duty_freq}
    \end{subfigure}
    
    \caption{(a) Stability diagram of the QMF operating at 61.20\% and 61.16\% duty cycles. (b) Stability diagram in duty cycle--frequency space ($a=0$) for $m/z = 40$, $r_{0} = 4\,\mathrm{mm}$, $V_{0} = 100\,\text{V}$.}
    \label{fig:combined-stability}
\end{figure*}

A typical $a-q$ stability diagram is illustrated in Fig.~\ref{fig:combined-stability}(a), showing a shrinking of the stability region with increasing duty cycle.  Fig.~\ref{fig:combined-stability}(b) represents the stability boundary in the $d$–$f$ parameter space and provides a direct guideline for selecting an appropriate duty cycle at a given operating frequency to achieve an optimal balance between resolution and transmission. 

Clearly, from Fig.~\ref{fig:combined-stability}(a), an increase in the duty cycle leads to an enhancement of the nominal resolution, defined as the ratio of the apex of the stability diagram to the separation between the points where the scan line intersects its boundaries. As reported by Simke et al.,\cite{simke2022simulations} a decrease in the duty cycle results in a narrower transmission profile accompanied by reduced transmission, in agreement with the stability diagram. In the present work, an intermediate duty cycle of $61.20\%$ is selected as an optimal compromise between resolution and transmission efficiency.

\begin{figure}
    \centering
    \includegraphics[width=0.48\linewidth]{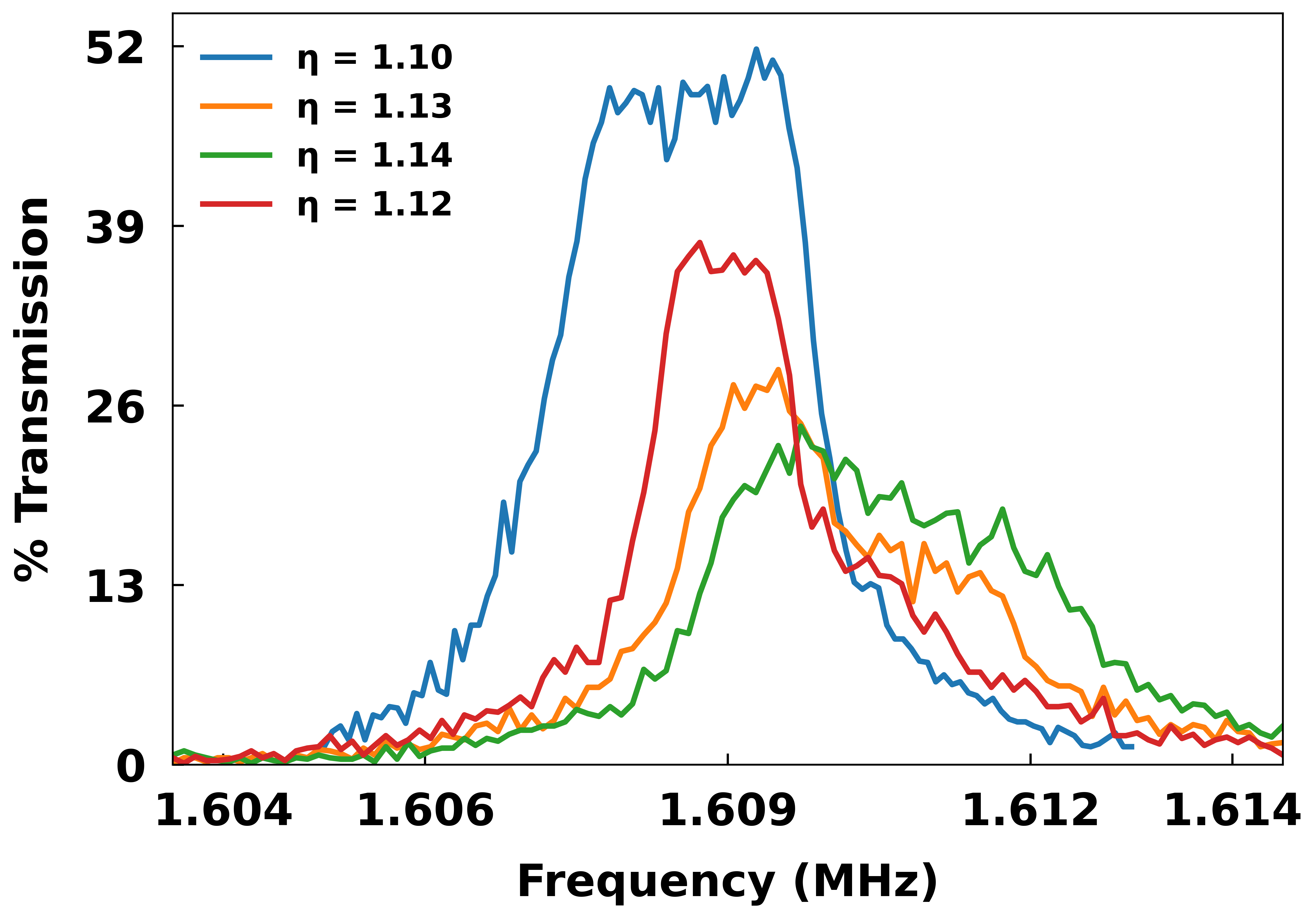}
    \caption{Transmission characteristics for symmetric setup with variation in the geometry parameter $\eta$.}
    \label{fig:duty_cycle_variation}
\end{figure}

A QMF with circular rods inherently generates higher-order multipole components, particularly the dodecapole component, in addition to the ideal quadrupole field. The strength of the dodecapole component depends on the rod-to-field radius ratio ($\eta=r/r_0$) and is known to influence QMF performance, as extensively studied for sinusoidal excitation~\cite{sysoev2022balance,douglas2002influence}. In the present study, in order to select an appropriate value of $\eta$, transmission simulations were carried out using a digital waveform with a duty cycle of $61.2\%$. While keeping $U=0$ and $V_0$ constant, the excitation frequency was varied at a fixed duty cycle. Fig.~\ref{fig:duty_cycle_variation} shows the transmission as a function of frequency for different values of $\eta$ at a duty cycle of $61.20\%$. It is observed that variations in $\eta$ affect both the width and intensity of the transmission profile. Based on this analysis, $\eta=1.12$ yields a narrower transmission peak and is therefore selected, together with a duty cycle of $61.20\%$ (and $U=0$), as an optimal compromise between transmission width and intensity for the remainder of this study. 

\section{Digital QMF: Asymmetric Geometry}
This section presents a systematic study of the transmission characteristics of a digital QMF in the presence of radial asymmetry. The asymmetry is introduced through variations in electrode radii and positional displacements, applied either to a single electrode or to a pair of diagonally opposite electrodes. Specifically, four cases are considered along the $x$-axis: (i) variation of the radii of both rods, (ii) displacement of both rods, (iii) variation of the radius of a single rod, and (iv) displacement of a single rod, as shown schematically in Fig.~1. The effect of initial state reversal of the digital waveform on transmission characteristics is also examined.

The degree of asymmetry is parameterized by $\gamma$, defined as $\gamma = \Delta x / r_0 $. Here $\Delta x$ represents either the displacement of an electrode along the $x$-axis from its symmetric position or the effective radial deviation defined as $\Delta x=r-r_a$, where $r_a$ is the radius of the electrode(s) along the $x$-axis and $r$ is the radius in the symmetric configuration.

For a fixed field radius $r_0$ and a given $\Delta x$, the degree of asymmetry ($\gamma$) is identical for both radius variation and rod displacement. However, since these two modes arise from different geometric modifications, they perturb the multipole field components differently and consequently lead to distinct transmission characteristics in the QMF, as reported in previous studies\cite{Jana2025,Jana2026}.

\subsection{Diagonal Asymmetry}
Radial asymmetry arising from variations in electrode radii or displacements of a diagonally opposite rod pair leads to unequal field dimensions ($x_0 \neq y_0$) along the transverse directions in the radial plane. In contrast to the symmetric geometry ($x_0 = y_0 = r_0$), the asymmetric configuration in this case results in field dimensions as $x_0 = r_0 + \Delta x$ and $y_0 = r_0$. To account for such geometric deviations, the radial potential model developed by our group\cite{Jana2026} is employed. Under this perturbation, the equations of motion can be recast in normalized form as
\begin{equation}\label{eq8}
\ddot{x} + \frac{a + 2q S(\xi)}{p}\,x = 0, \quad 
\ddot{y} - \frac{a + 2q S(\xi)}{p}\,y = 0,
\end{equation}
where
\begin{equation}
p = \frac{1 + \left(1 + \gamma\right)^2}{2}
\end{equation}
is a geometric correction factor that accounts for the radial asymmetry of the electrode configuration \cite{Jana2026}.

The stability diagram for the case of diagonal electrode-pair asymmetry is obtained using the transfer matrix formalism within Floquet theory, incorporating the correction factor $p$. The modified transfer matrix is then employed to compute the corresponding shift in the stability regions. This approach enables the systematic analysis of the modified first stability region associated with the quadrupole component.

\begin{figure}
    \centering
    \includegraphics[width=1.0\linewidth]{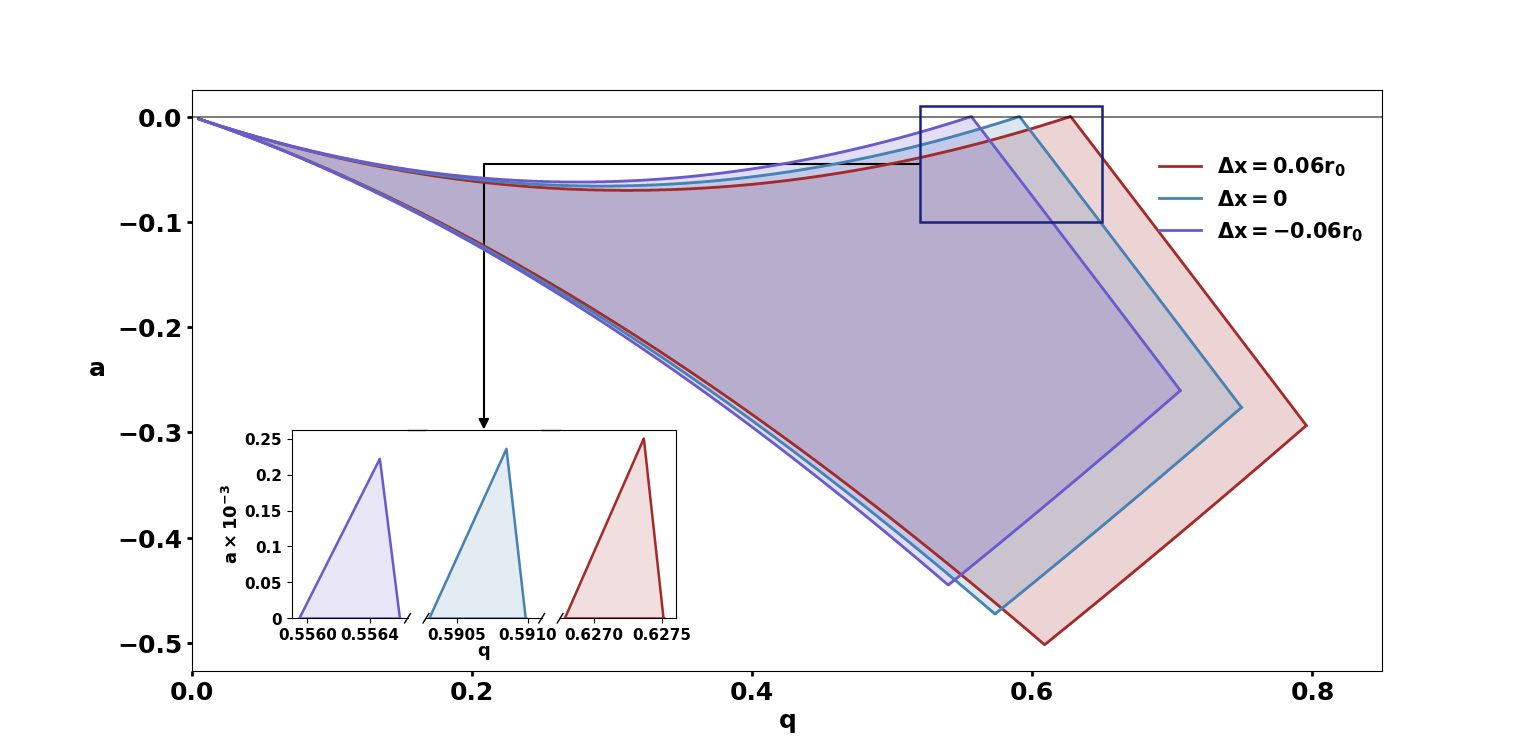}
    \caption{First stability zone for $\gamma = -0.06, 0$ and $+0.06$, illustrating the shift of the stability apex with radial asymmetry at a constant duty cycle of $61.20\%$.}
    \label{fig:sd-shift}
\end{figure}

Fig.~\ref{fig:sd-shift} illustrates the stability diagrams for the symmetric ($\gamma = 0$) and asymmetric ($\gamma = \pm 0.06$) configurations. The introduction of radial asymmetry distorts and shifts the stability boundaries: an expansion in the field dimension ($\Delta x > 0$) shifts the apex towards higher $q$ values, while a contraction ($\Delta x < 0$) shifts it towards lower $q$ values. Since $q \propto 1/f^2$, these shifts directly translate into corresponding changes in the operating frequency—higher $q$ implying lower frequency and vice versa. These modifications alter the effective operating conditions and are expected to influence the transmission characteristics of the QMF.

\begin{figure*}[htbp]
    \centering
    \begin{subfigure}[b]{0.47\textwidth}
        \centering
        \setlength{\unitlength}{1mm}
        \begin{picture}(0,0)
            \put(-35, 2){\textbf{(a)}} 
        \end{picture}
        \includegraphics[width=\textwidth]{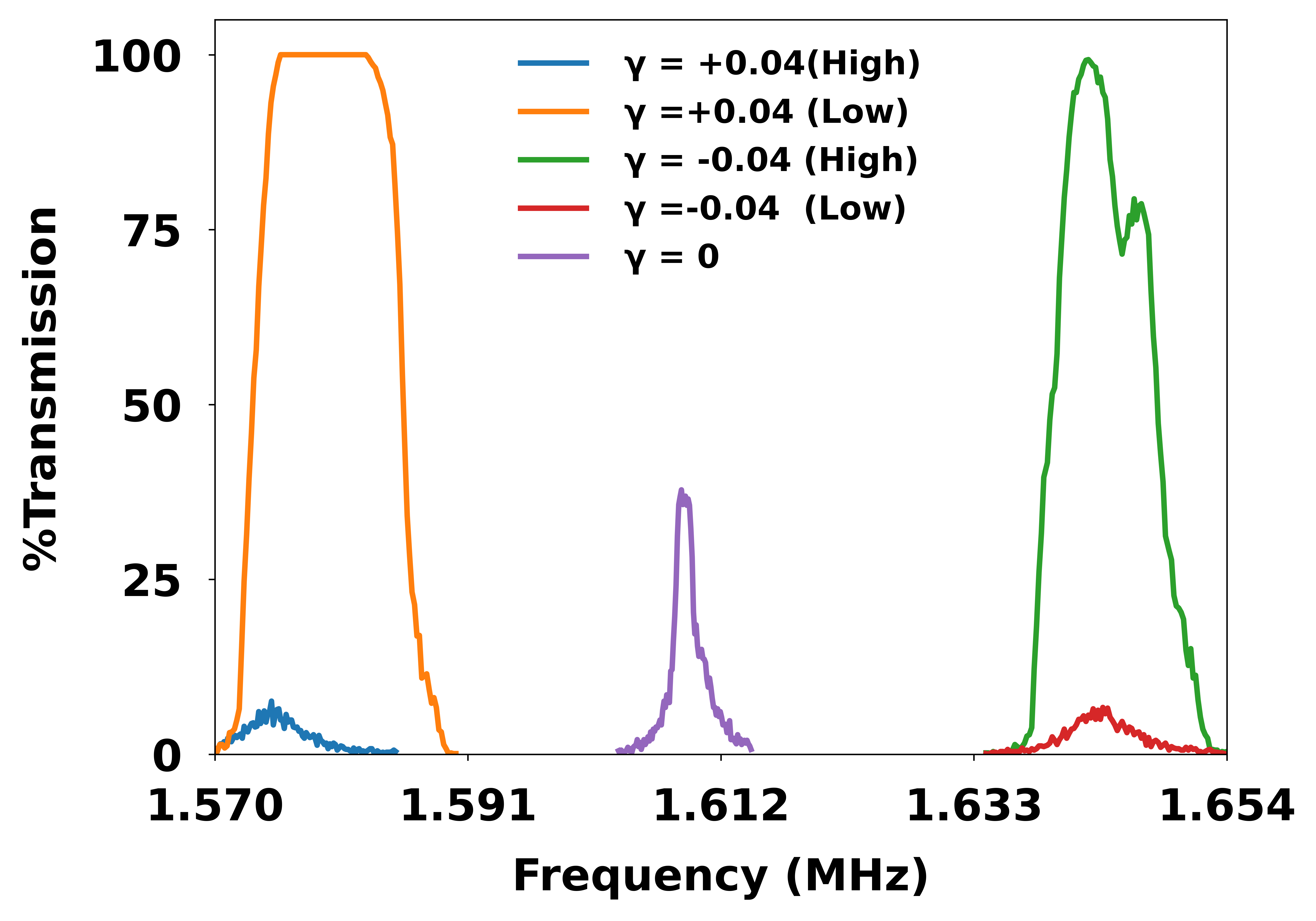}
        \label{fig:double-rod-radius-transmission}
    \end{subfigure}
    \hfill
    \begin{subfigure}[b]{0.47\textwidth}
        \centering
        \setlength{\unitlength}{1mm}
        \begin{picture}(0,0)
            \put(-35, 2){\textbf{(b)}}
        \end{picture}
        \includegraphics[width=\textwidth]{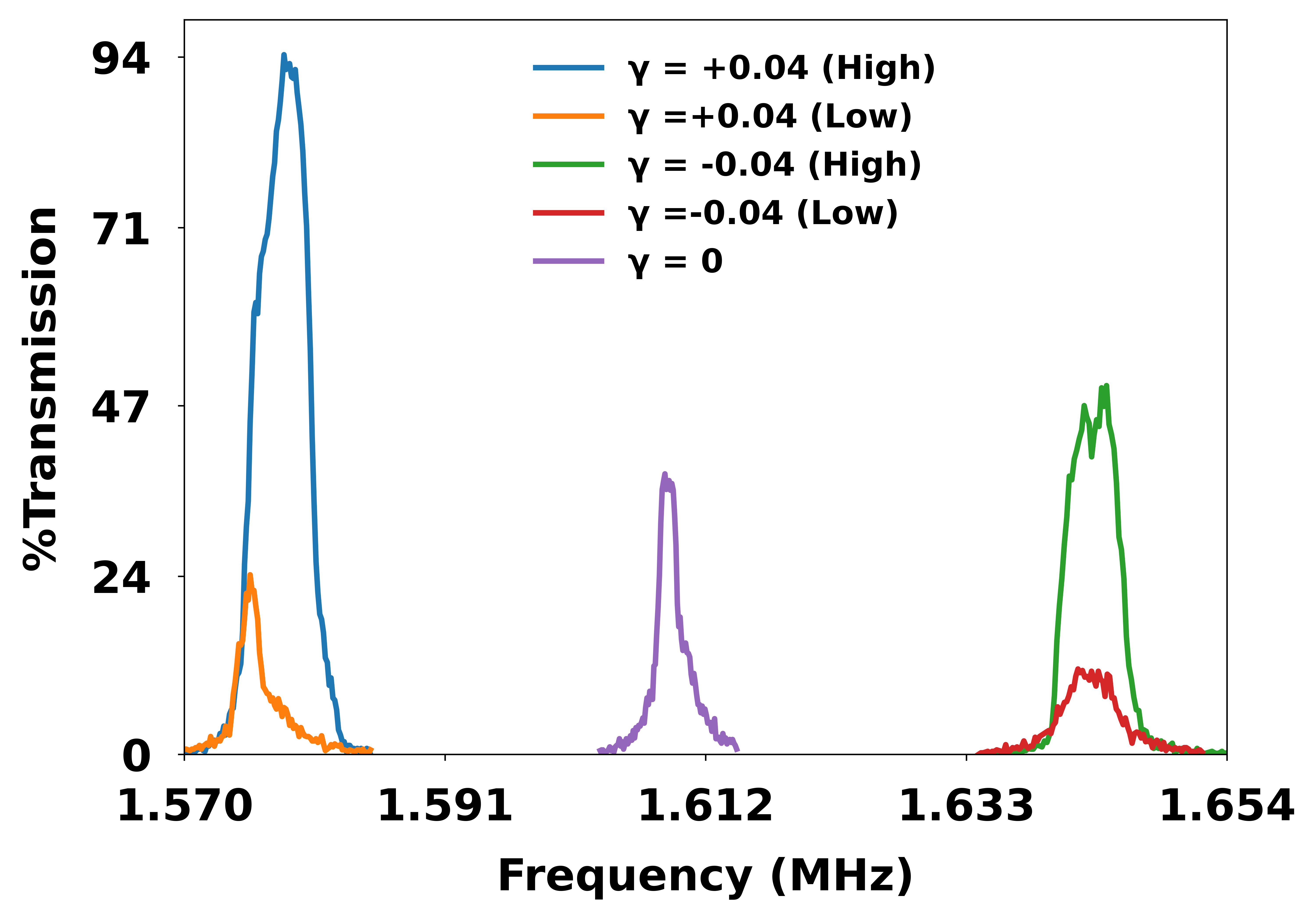}
        \label{fig:dis_doublerod}
    \end{subfigure}
    
    \caption{Transmission characteristics of the QMF with diagonal asymmetry ($\gamma=\pm 0.04$): (a) radius variation and (b) electrode displacement, with the asymmetric electrode pair initially set to the high and low states of the applied voltage.}
    \label{fig:double-rod-combined}
\end{figure*}
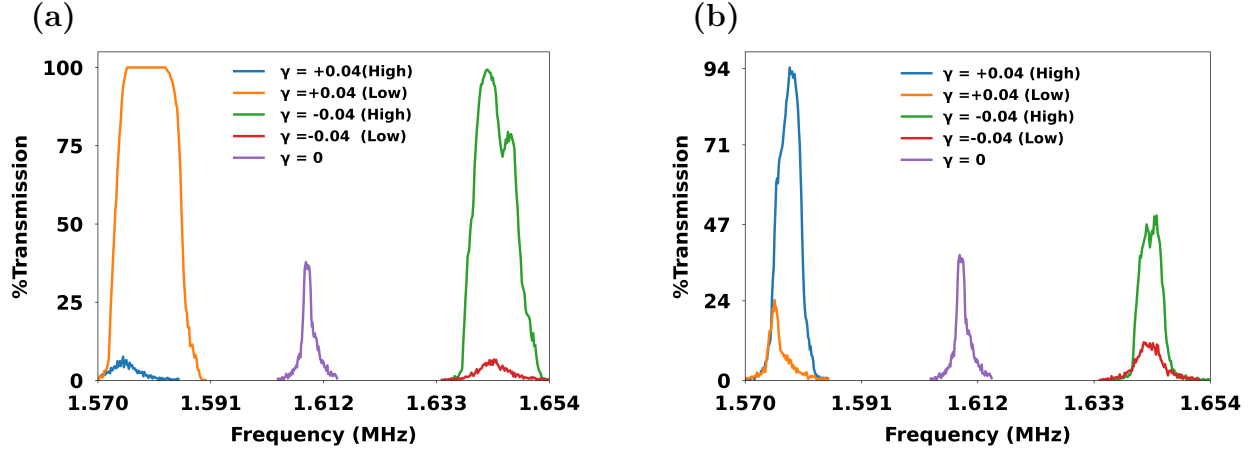
Fig.~\ref{fig:double-rod-combined} shows the transmission as a function of frequency for diagonal electrode-pair asymmetry introduced via radius variation (panel~(a)) and displacement (panel~(b)), for $\gamma = 0$ and $\gamma = \pm 0.04$, with the asymmetric pair initially subjected to high ($+V_0$) and low ($-V_0$) RF pulse.

As shown in Fig.~5(a), a narrow transmission peak appears at approximately $1.61$~MHz in the symmetric configuration ($\gamma=0$). In the presence of asymmetry introduced via radius variation of both rods, the transmission shifts in frequency space: an increased-radius configuration ($\gamma = -0.04$) shifts the peak towards higher frequencies, whereas a decreased-radius configuration ($\gamma = 0.04$) shifts it towards lower frequencies. This trend is systematic with $\gamma$ and is consistent with the shift in the stability apex in the radially asymmetric QMF~(Fig.~\ref{fig:sd-shift}) discussed in the previous section. A distinct feature is observed upon reversal of the initial waveform state: for $\gamma = 0.04$, the transmission increases significantly, whereas for $\gamma = -0.04$, it decreases when the asymmetric rod pair is initially subjected to the low state of the RF pulse (i.e., $-V_0$).

A similar trend is observed for diagonal displacement asymmetry (Fig.~5(b)). Increasing asymmetry leads to a systematic shift in the transmission peak along the frequency axis and a pronounced dependence on the initial waveform state, affecting both peak position and intensity.

\subsection{Single Rod Asymmetry}

Following the analysis of diagonal asymmetry, we now consider the limiting case where asymmetry is introduced in a single electrode. In this configuration, radial imperfections are implemented either by varying the radius of a single electrode along the $x$-axis or by displacing it along the same axis. Unlike diagonal asymmetry, where transverse symmetry about the center is preserved, this perturbation breaks the geometric symmetry of the system, resulting in a strongly distorted field that cannot be modeled using the earlier approach. Consequently, the stability regions cannot be constructed within the standard Floquet framework, and the QMF performance is analyzed directly in terms of transmission characteristics for a fixed duty cycle of $61.20$\%.

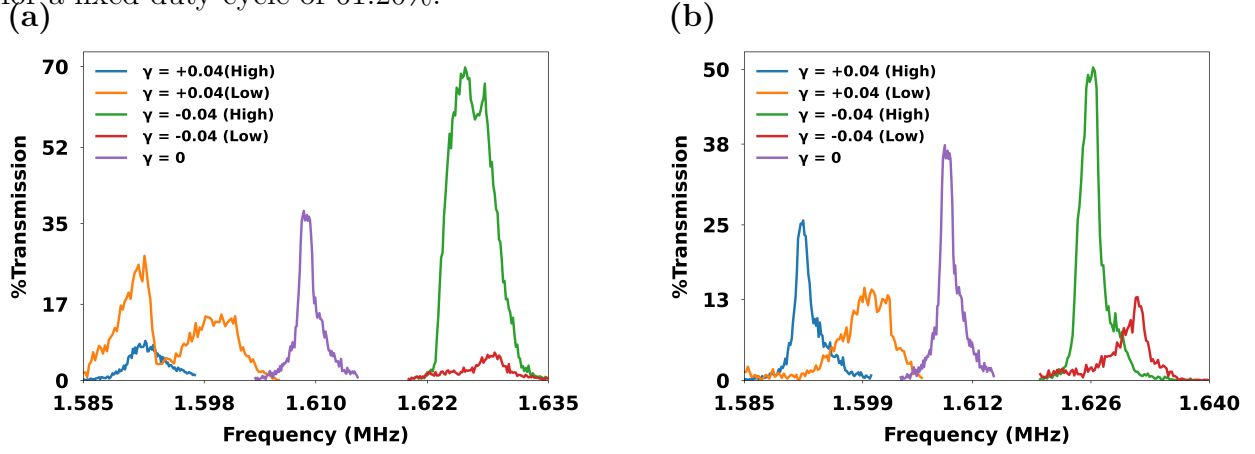
\begin{figure*}[htbp]
    \centering
    \begin{subfigure}[b]{0.47\textwidth}
        \centering
        \setlength{\unitlength}{1mm}
        \begin{picture}(0,0)
            \put(-38, 2){\textbf{(a)}} 
        \end{picture}
        \includegraphics[width=\textwidth]{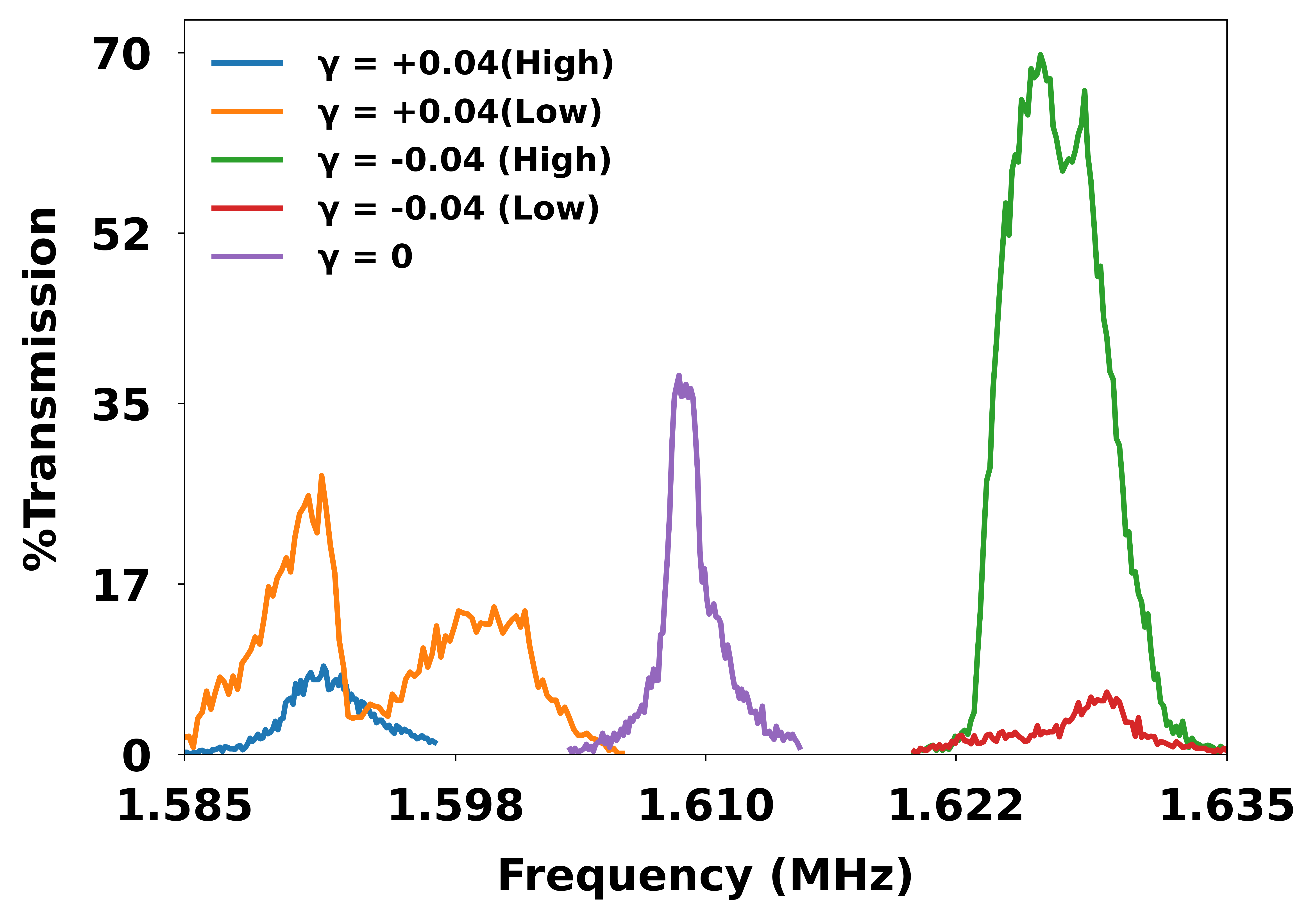}
        \label{fig:singlerod_radius}
    \end{subfigure}
    \hfill
    \begin{subfigure}[b]{0.47\textwidth}
        \centering
        \setlength{\unitlength}{1mm}
        \begin{picture}(0,0)
            \put(-38, 2){\textbf{(b)}}
        \end{picture}
        \includegraphics[width=\textwidth]{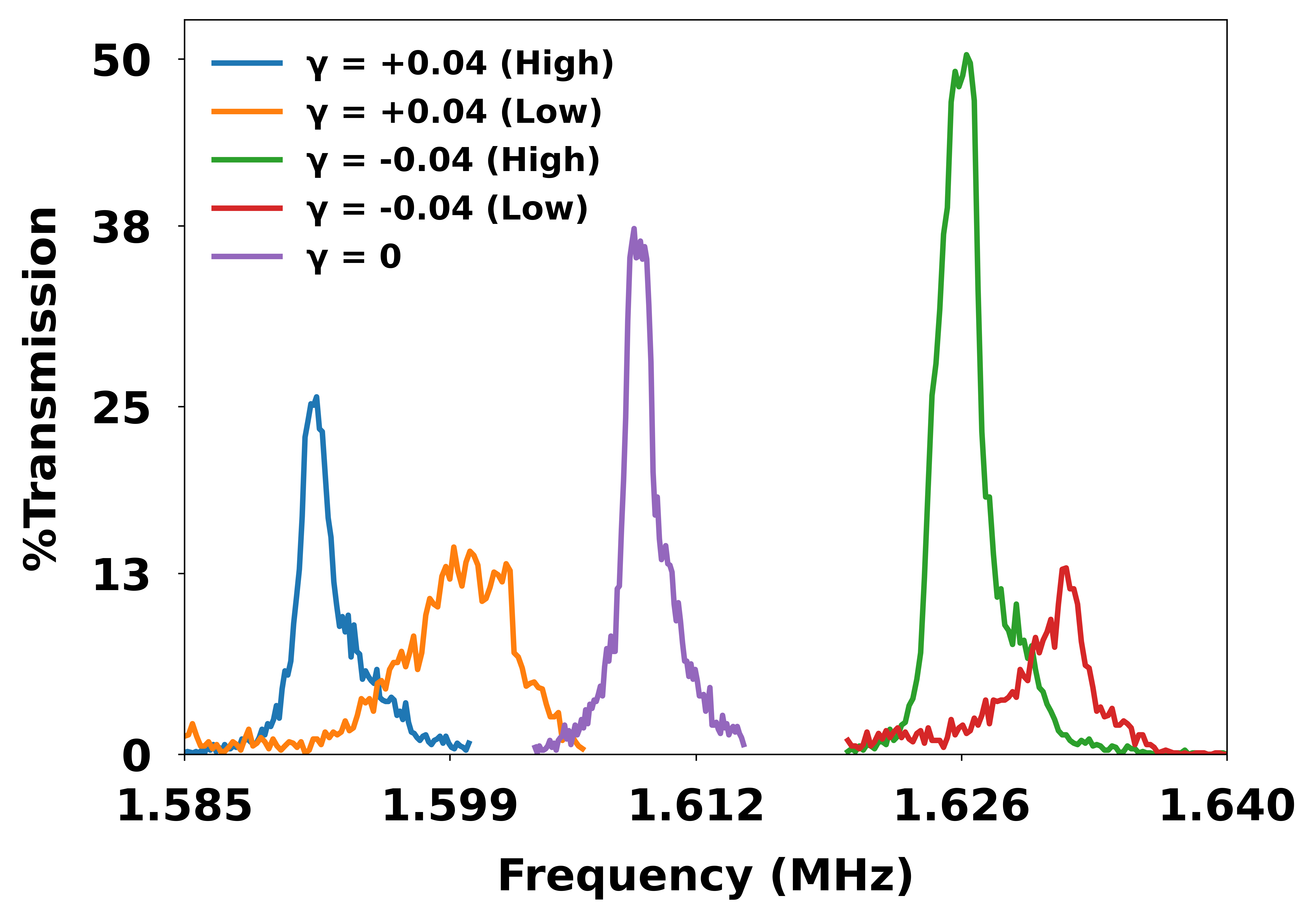}
        \label{fig:singlerod_displacement}
    \end{subfigure}
    
    \caption{Transmission characteristics of the QMF with single-rod asymmetry ($\gamma=\pm 0.04$): (a) radius variation and (b) electrode displacement, with the asymmetric electrode and its diagonal counterpart initially set to the high and low states of the applied voltage.}
    \label{fig:single-rod-combined}
\end{figure*}

Fig.~\ref{fig:single-rod-combined} shows the transmission characteristics for single-rod radial asymmetry at $\gamma = 0$ and $\gamma = \pm 0.04$, for both radius variation and electrode displacement, with the initial waveform state set to high ($+V_0$) and low ($-V_0$) on the odd electrode and its diagonal counterpart. Qualitatively similar trends to those observed for diagonal asymmetry are obtained: the transmission width and intensity exhibit a pronounced dependence on both the degree of asymmetry and the initial RF state. Notably, a distinct precursor peak emerges, particularly for $\gamma = 0.04$ in the case of single-rod radius variation under an initial low waveform state~(Fig.~6(a)).

\section{Discussion}
For a quantitative comparison of QMF performance at different degrees of asymmetry, the resolution $R$ is extracted from the corresponding transmission characteristics. In general, the resolution is defined as $R = m/\Delta m$, where $m$ is the peak mass and $\Delta m$ is the peak width, typically measured as full width at half maximum (FWHM). For rectangular wave operation, the resolution can equivalently be expressed as $R = f/2\Delta f$, where $f$ and $\Delta f$ denote the peak frequency and the corresponding FWHM of the transmission curve, respectively\cite{simke2022simulations}.

\begin{figure*}[htbp]
    \centering
    \begin{subfigure}[b]{0.47\textwidth}
        \centering
        \setlength{\unitlength}{1mm}
        \begin{picture}(0,0)
            \put(-45, -1){\textbf{(a)}} 
        \end{picture}
        \includegraphics[width=\textwidth]{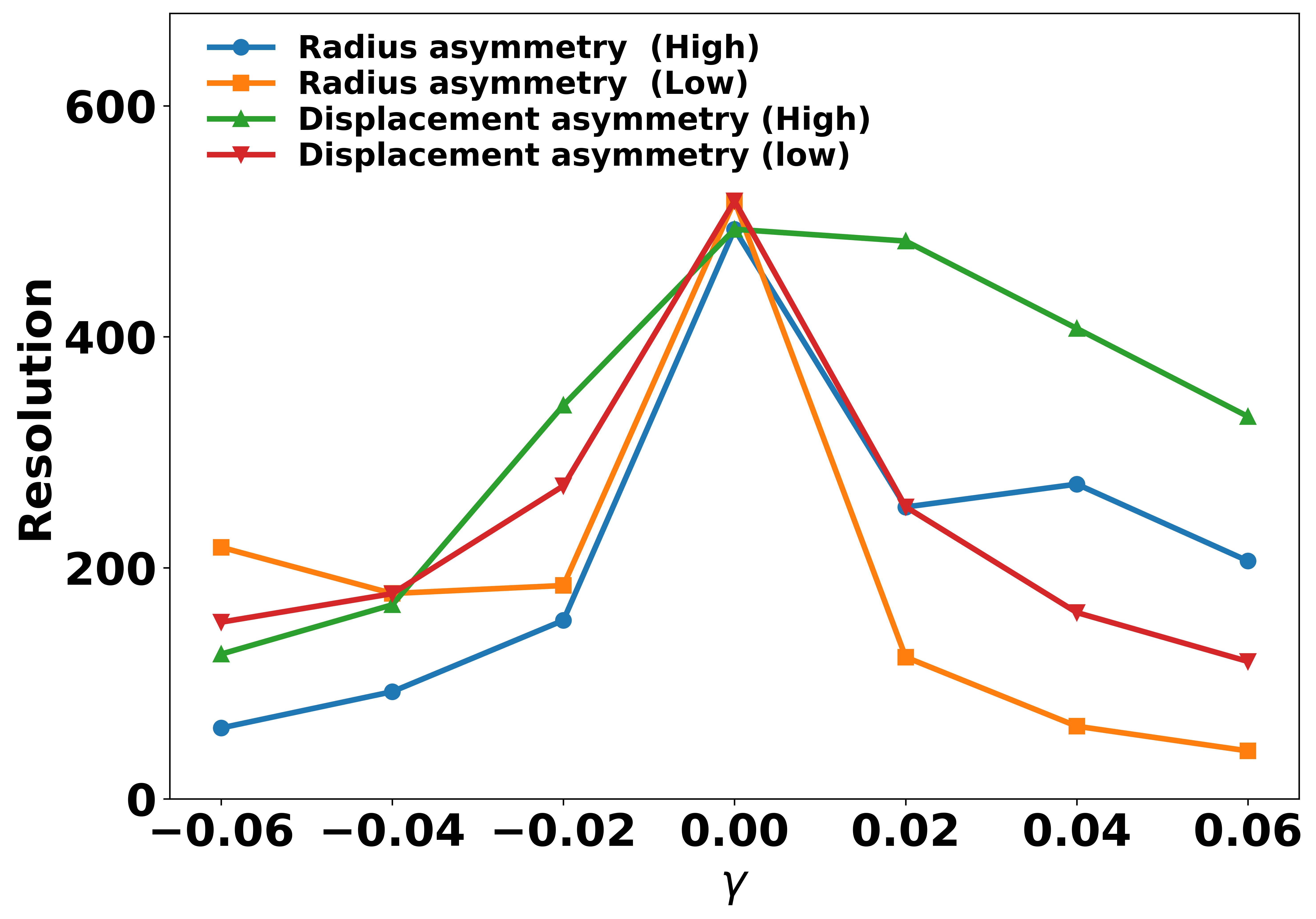}
        \label{fig:R-vs-gamma-all-diagonal}
    \end{subfigure}
    \hfill
    \begin{subfigure}[b]{0.47\textwidth}
        \centering
        \setlength{\unitlength}{1mm}
        \begin{picture}(0,0)
            \put(-45, -1){\textbf{(b)}}
        \end{picture}
        \includegraphics[width=\textwidth]{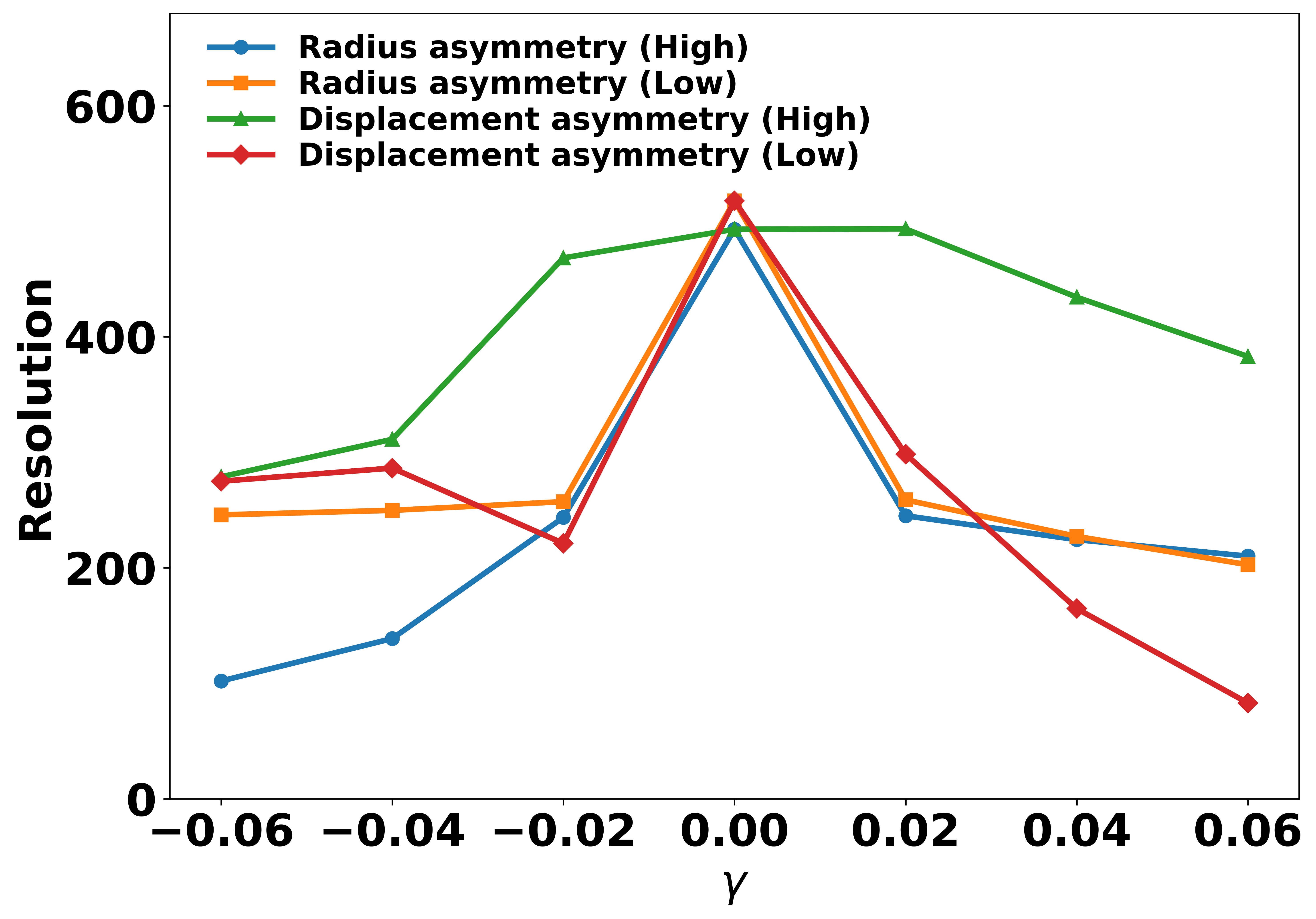}
        \label{fig:R-vs-gamma-all-single}
    \end{subfigure}
    
    \caption{Variation of resolution with the asymmetry parameter ($\gamma$) for (a) diagonal asymmetry and (b) single-rod asymmetry. The high and low states indicated in the legends correspond to the initial state of the RF voltage applied to the asymmetric rod pair.}
    \label{fig:R-vs-gamma-combined}
\end{figure*}
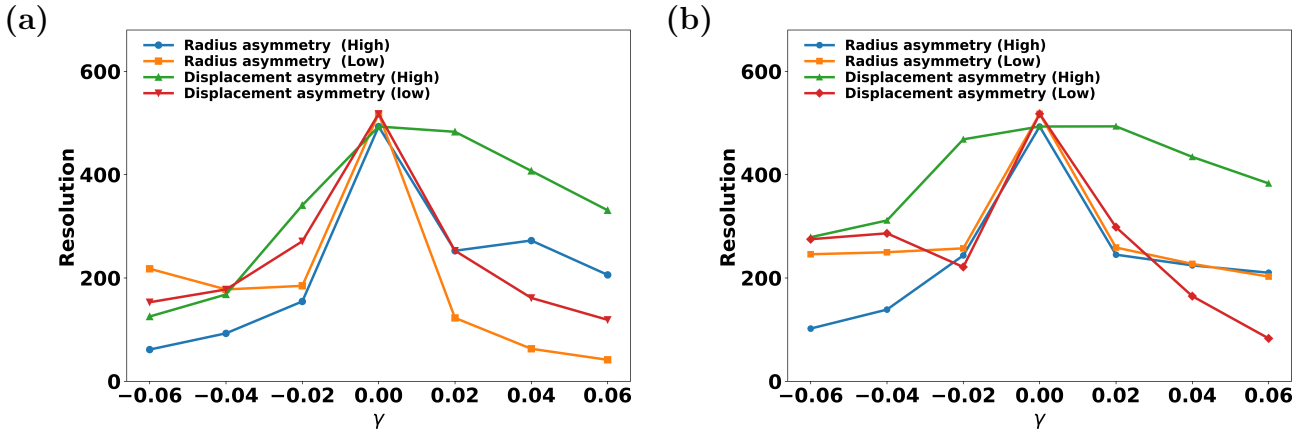

The resolution $R$, extracted from the transmission characteristics, is plotted in Fig.~\ref{fig:R-vs-gamma-combined}(a) as a function of $\gamma$ for diagonal asymmetry introduced via both radius variation and electrode displacement. In all cases, $R$ decreases with increasing asymmetry parameter $|\gamma|$, irrespective of the initial state of the applied RF pulse, confirming that the symmetric configuration yields optimal performance.

The extent and trend of resolution degradation depend strongly on the type and degree of asymmetry, as well as the initial waveform state applied to the $x$-pair electrodes. The displacement asymmetry exhibits a more gradual reduction in $R$ when the displaced electrode pair is biased at an initial high waveform state ($+V_0$), particularly for $\gamma > 0$, indicating greater tolerance to such imperfections. Upon reversal of the initial waveform state, however, a steeper decline in $R$ is observed for both types of diagonal asymmetry.

Fig.~\ref{fig:R-vs-gamma-combined}(b) shows the resolution extracted from the transmission characteristics for single-rod asymmetry, introduced via both radius variation and electrode displacement, with the initial waveform state set to low ($-V_0$) and high ($+V_0$) for the $x$-pair electrodes. A trend similar to that observed for diagonal asymmetry is obtained: the resolution decreases with increasing asymmetry parameter $|\gamma|$ in all cases. However, displacement asymmetry exhibits a relatively slower reduction in $R$ over a broader range of $\gamma$ for an initial high waveform state, indicating greater tolerance to such imperfections.

As reported in previous studies, radial asymmetry in a quadrupole mass filter introduces a significant octupole component in addition to the intrinsic dodecapole component associated with circular rod geometry\cite{Jana2025,Jana2026,dutta2026influence}. The strength of the octupole component increases with the degree of asymmetry and generally leads to a degradation in resolution for QMFs driven by sinusoidal voltages. The observed reduction in resolution in a radially asymmetric digital QMF can thus be attributed to the combined influence of higher-order multipole fields, particularly the octupole and dodecapole components. Furthermore, the presence of the octupole component has been shown to induce a polarity dependence of the resolution with respect to the applied DC voltage. In a digital QMF operated at duty cycles other than $50\%$, an effective static voltage component arises; consequently, in the presence of radial asymmetry, the QMF performance becomes sensitive to the initial state of the RF pulse applied to the odd electrodes.

To investigate the origin of precursor formation in the case of single-rod radius variation asymmetry (Fig.~\ref{fig:single-rod-combined} a)), the transmission characteristics are analyzed for $\gamma = 0.04$ under an initial low waveform state, where the precursor effect is most pronounced, and compared with the case of an initial high waveform state. The role of higher-order multipole components and state reversal is examined through SIMION simulations of the stability boundaries, following the methodology developed by our group and described in \cite{dutta2026influence}. In this approach, the stability diagram is constructed via a two-dimensional grid scan in the $a$–$q$ space.

\begin{figure*}[htbp]
    \centering
    \begin{subfigure}[b]{0.47\textwidth}
        \centering
        \setlength{\unitlength}{1mm}
        \begin{picture}(0,0)
            \put(-35, 1){\textbf{(a)}} 
        \end{picture}
        \includegraphics[width=\textwidth]{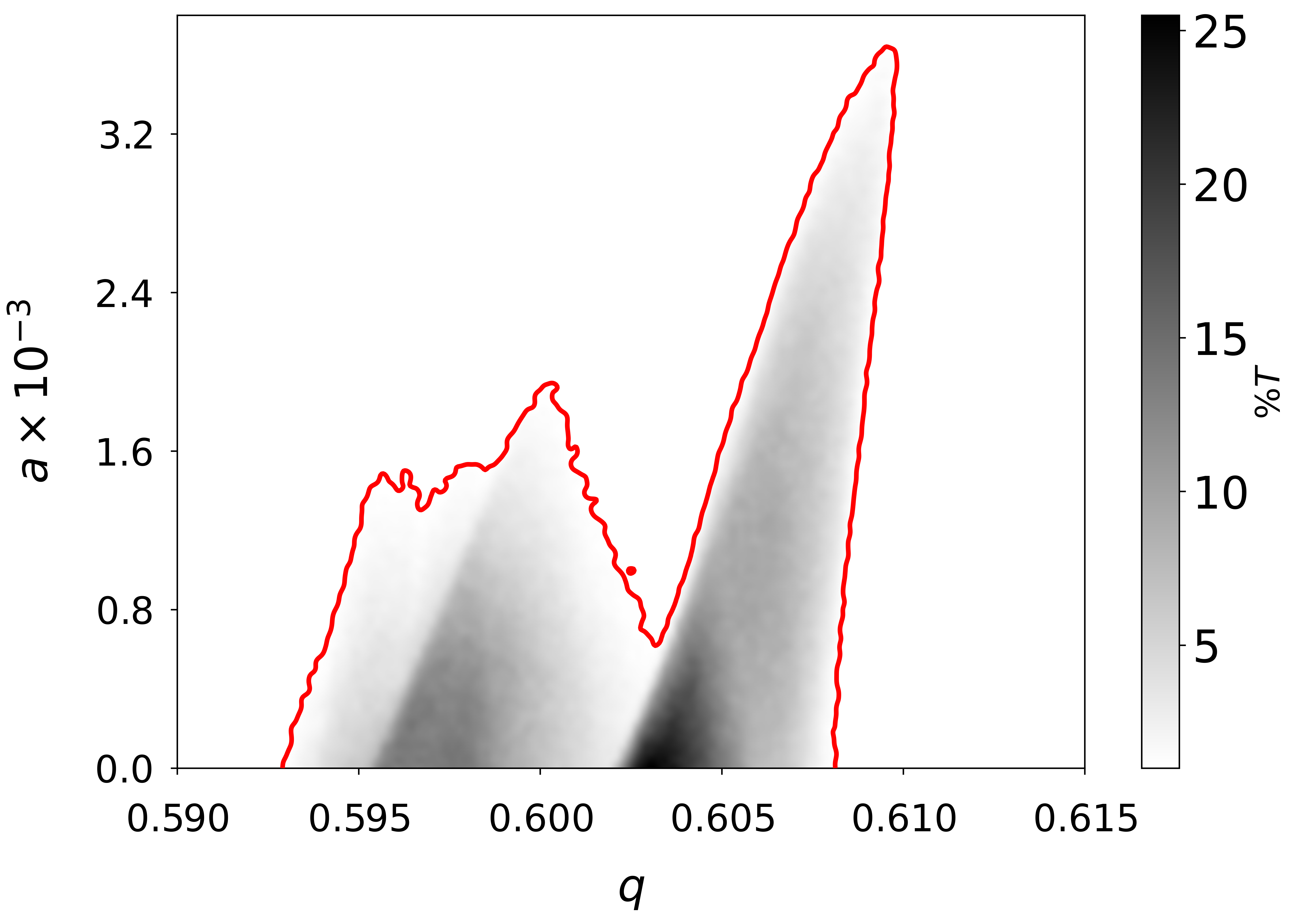}
        \label{fig:+0.04_radius_polarity}
    \end{subfigure}
    \hfill
    \begin{subfigure}[b]{0.47\textwidth}
        \centering
        \setlength{\unitlength}{1mm}
        \begin{picture}(0,0)
            \put(-35, 1){\textbf{(b)}}
        \end{picture}
        \includegraphics[width=\textwidth]{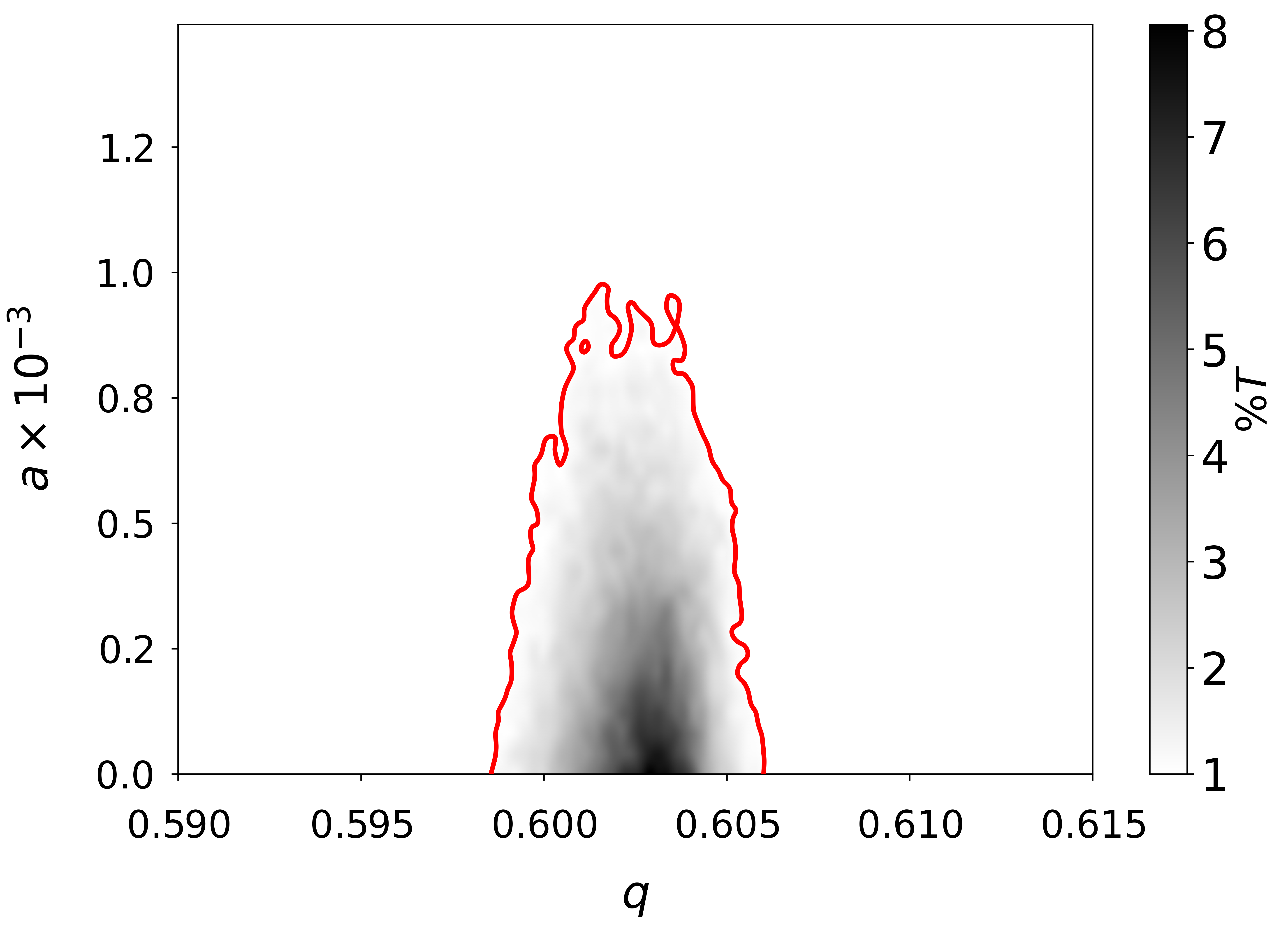}
        \label{fig:+0.04_radius_normal}
    \end{subfigure}
    
    \caption{SIMION stability diagrams for 5000 particles and 10000 points at $\gamma = 0.04$ for single-rod radius variation asymmetry: (a) reverse polarity and (b) normal polarity. The stability boundary is defined at 1\% transmission in both cases.}
    \label{fig:simion_04}
\end{figure*}
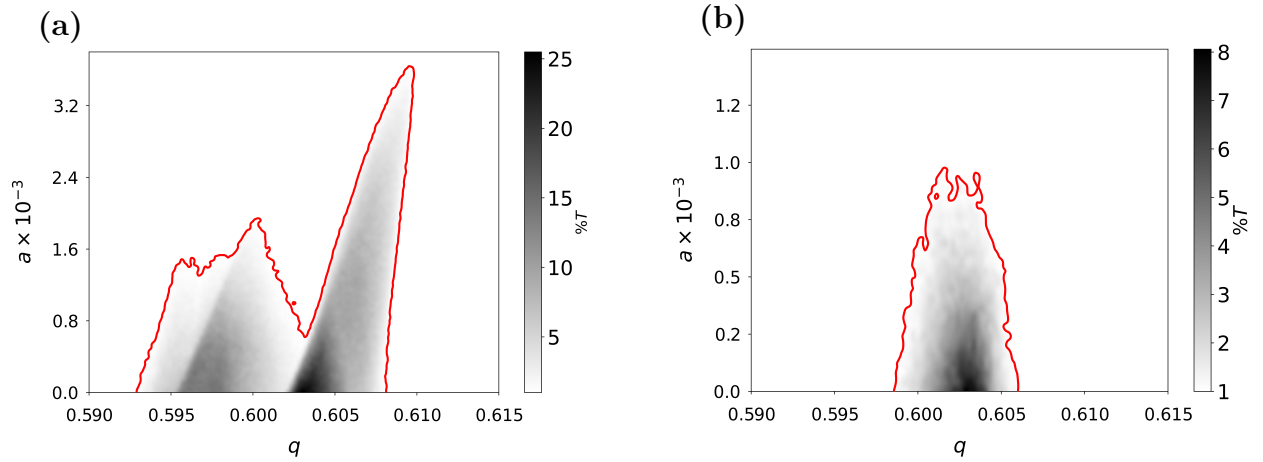

The stability diagrams obtained for initial high and low waveform states at $\gamma = 0.04$ are shown in Fig.~\ref{fig:simion_04}. A striking feature is observed in the initial state-reversed case, where a clear bifurcation appears, likely originating from a nonlinear resonance. When mapped to the frequency domain, this bifurcation corresponds to the same frequency range over which the transmission contour exhibits splitting. This correlation suggests that the splitting observed in the stability diagram provides a plausible explanation for the emergence of precursor peaks in the transmission profile. In contrast, no such bifurcation is observed for the normal polarity, consistent with the absence of precursor peaks.

\section{Conclusion}

In this work, we have investigated the performance of a rectangular-wave driven quadrupole mass filter in the presence of controlled radial asymmetry, introduced as an effective octupolar perturbation. Four distinct configurations are considered, namely single rod radius variation, single rod displacement, diagonal rod-pair radius variation, and diagonal rod-pair displacement, enabling a systematic assessment of geometry-induced field distortions under digital excitation. 

Across all configurations, radial asymmetry is found to degrade resolution, establishing a consistent and general impact of octupolar contributions in rectangular-wave operated QMFs. A key result of the present study is the pronounced dependence of transmission characteristics on the initial state of the RF pulse applied at the asymmetric pair of rods. This state sensitivity is observed in all cases and significantly influences peak shape, transmission efficiency, and the extent of asymmetry-induced distortions.

Notably, under single rod radius variation, precursor (satellite) peaks emerge for specific initial waveform state. Stability analysis indicates that these features originate from a bifurcation of the stability region, linking the observed transmission anomalies directly to modifications in the underlying stability diagram, indicating role of higher order spatial harmonics. 

The results demonstrate that while rectangular wave excitation provides additional tunability, it also amplifies sensitivity to geometric imperfections and phase conditions. These findings establish clear design and operational constraints for digital QMFs and provide a framework for understanding and mitigating asymmetry-induced performance degradation in high-resolution mass filtering applications.

\begin{acknowledgement}
ND thanks SERB/ANRF India (CRG/2023/001529) and BRNS India (58/14/21/2023 - BRNS12329) for funding. 
\end{acknowledgement}

\section{Conflicts of Interest}
The authors declare no conflicts of interest for this manuscript.
\bibliography{reference}

\end{document}